\documentclass[a4paper,11pt]{article}

\usepackage{pos}
\usepackage{lipsum} %package to generate placeholder text in the following

\title{All-flavor Time-dependent Search for Transient Neutrino Sources}

\ShortTitle{All-flavor Time-dependent Search for Transient Neutrino Sources}

\author{The IceCube Collaboration \\{\normalsize \normalfont(a complete list of authors can be found at the end of the proceedings)}\\}

\emailAdd{jcarpio@icecube.wisc.edu}
\emailAdd{ali.kheirandish@icecube.wisc.edu}
\emailAdd{hmniederhausen@icecube.wisc.edu}

\abstract{

Transient sources are among the preferred candidates for the sources of high-energy neutrino emission. Intriguing examples so far include blazar flares and tidal disruption events coincident with IceCube neutrinos. Here, we report the first all-flavor, all-sky time-dependent search for neutrino sources by combining IceCube throughgoing tracks, starting tracks and cascades. Throughgoing tracks provide the best sensitivity in the Northern Sky, while cascades have worse angular resolution but yield better sensitivity in the Southern Sky than tracks. The relatively new starting tracks sample has reduced contamination from atmospheric muons. This analysis takes advantage of the strengths of each of the datasets, combining them for increased statistics and obtaining the best accessible all-sky sensitivity for transient searches. In this search, we look for unbound $E^{-\gamma}$ power-law sources, as well as $E^{-2}$ sources with low and high-energy exponential cutoffs, optimizing the sensitivity for the duration of the flares. 
\vspace{4mm}

{\bfseries Corresponding authors:}
Jose Carpio$^{1*}$, 
Ali Kheirandish$^{1}$,
Hans Niederhausen$^{2}$\\

{$^{1}$ \itshape University of Nevada Las Vegas}\\
$^{2}$ \itshape Michigan State University\\[4mm]
$^*$ Presenter
}

\ConferenceLogo{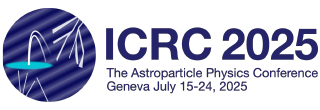}

\FullConference{39th International Cosmic Ray Conference (ICRC2025)\\
 15–24 July 2025\\
Geneva, Switzerland\\}

\begin{document}

\maketitle

\section{Introduction}
Transient neutrino sources, such as blazar flares, tidal disruption events, and gamma-ray bursts, are among the primary candidates for the origin of high-energy neutrino emission~\cite{Murase:2019tjj}. 
In 2017, IceCube detected a 290 TeV neutrino spatially coincident with the blazar TXS 0506+056 ~\cite{IceCube:2018cha}. 
Time-integrated searches with 10 years of neutrino tracks showed the transient TXS 0506+056 as the second most significant source in the Northern Sky \cite{IceCube:2019cia}. 
However, the neutrino signal is not necessarily accompanied by an electromagnetic counterpart, leaving neutrinos as the only way to observe these sources. These would appear in neutrino data as a clustering of events in the time domain. 

Time-dependent searches provide an advantage over the time-integrated ones, for flares that last for $\Delta T\lesssim 100$ days, as it reduces the atmospheric neutrino background, which is relatively uniform in time. Previously, the IceCube collaboration has performed a transient search \cite{IceCube:2023myz} and multi-flare search \cite{IceCube:2021slf}, but did not find any new significant excesses.

In this study, we aim for improving upon previous searches by utilizing all channels of observations to obtain the best accessible sensitivity for IceCube.

Furthermore, neutrino point source searches have usually looked for generic power-law fluxes $E^{-\gamma}$.
However, the expected neutrino flux generally has features beyond a simple power-law. The IceCube collaboration has used alternative parametrizations for the diffuse neutrino flux (see e.g., \cite{Naab:2023xcz}). More recently, point source searches have also used more sophisticated signal energy spectra, such as Seyfert flux models to search for neutrino emission from Seyfert galaxies \cite{IceCube:2024dou}. Here, we implement a power-law with low and high-energy cutoffs  in the search that could capture the predicted features in prominent transient sources of neutrinos.

\section{Method}

For the analysis we will use the likelihood (LLH) function $\mathcal{L}$ \cite{Braun:2009wp}
\begin{equation}
\ln\mathcal{L}=\sum_i \ln\left[\mathcal{S}_i\frac{n_s}{N} + \left(1-\frac{n_s}{N}\right)\mathcal{B}_i\right],
\end{equation}
where $\mathcal{S}_i$ and $\mathcal{B}_i$ are the signal and background \textcolor{black}{probability density functions (pdfs)}, respectively, for each event $i$. 
The signal pdf can be factorized into a spatial pdf term, which describes the spatial clustering of the events around the source,  an energy pdf term, describing their energy distribution, and a temporal pdf term, describing the arrival time distribution of the events. 

This time-dependent analysis assumes that the temporal pdf follows a Gaussian distribution, centered at $T_0$ and a standard deviation $\sigma_T$. The spatial pdfs are two-dimensional Gaussians, where the reconstructed direction has a standard deviation $\sigma_i$. The NT sample was later enhanced by using a Kernel Density Estimate (KDE) so that the spectral shape of the flux is also taken into account, making the spatial pdf more realistic \textcolor{black}{(see e.g., supplemental material in \cite{IceCube:2022der} for a description of the KDE method)}.

The energy pdf is the probability that an event is observed with a given energy at a fixed declination $\delta$ and for a given spectral hypothesis. In the case of $E^{-\gamma}$ power-law sources, this introduces an additional parameter, the power-law index $\gamma$. 

\begin{figure*}
    
    \includegraphics[width=0.5\linewidth]{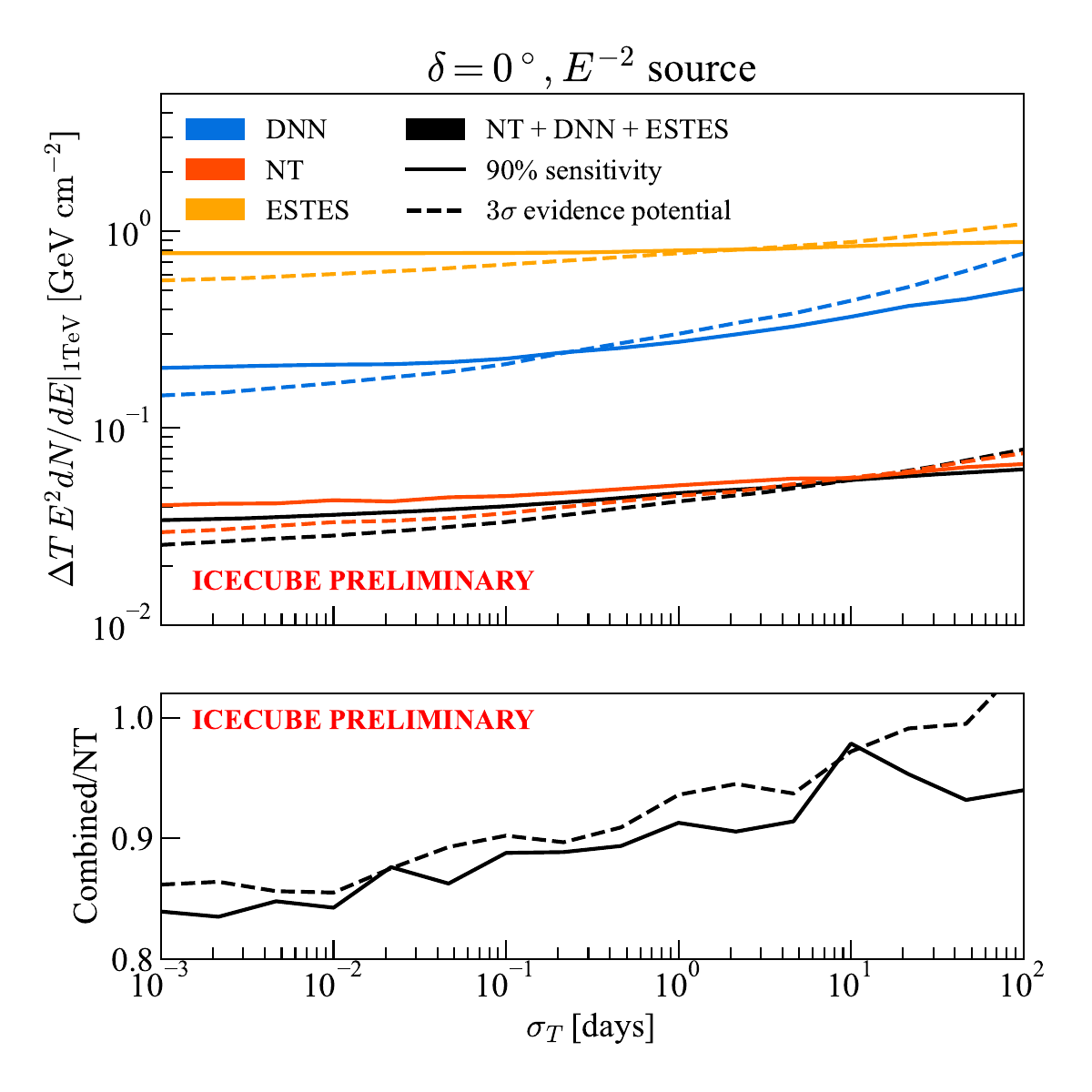}
    \includegraphics[width=0.5\linewidth]{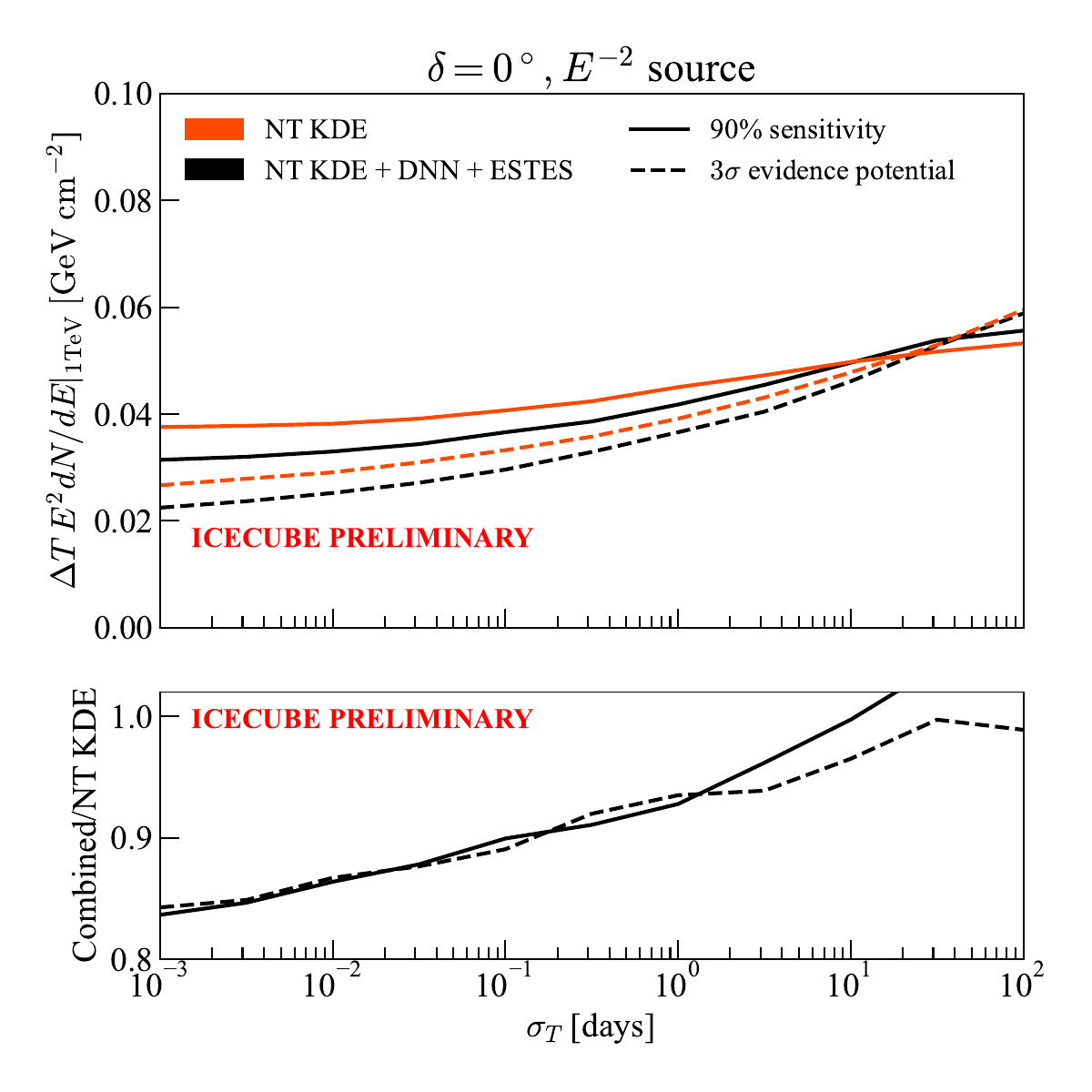}
    \caption{Top left panel: \textcolor{black}{Per-flavor} 90\% CL sensitivity (solid lines) and $3\sigma$ evidence (dashed lines) \textcolor{black}{fluences} for an $E^{-2}$ source at declination $\delta=0^\circ$, for different flare half-widths $\sigma_T$ and $E^{-\gamma}$ signal flux hypothesis. The blue, red, orange and black lines correspond to DNNCascade, NT, ESTES and combined datasets, respectively. Here, the NT sample assumes Gaussian spatial pdfs. Top right panel: Same as top left panel, but the NT dataset is using KDEs in its spatial and energy pdfs. Bottom left panel: Combined-to-NT fluence ratios. \textcolor{black}{Bottom right panel: Combined-to-NT KDE fluence ratios}}
    \label{Figure1}
\end{figure*}

Using $\mathcal{L}$, we define the test statistic

\begin{equation}
    {\rm TS} = 2\ln \left[\frac{\hat\sigma_T}{T_{\rm live}} \frac{\mathcal{L}(\hat n_s,\hat\gamma,\hat T_0,\hat\sigma_T)}{\mathcal{L}(n_s=0)}\right],
    \label{TSDefinition}
\end{equation}
where we used the power-law signal hypothesis as an example. $(\hat n_s,\hat\gamma,\hat T_0,\hat\sigma_T)$ is the parameter set that maximizes the term in square brackets. $T_{\rm live}$ is the livetime of the dataset. The term $/\sigma_T/T_{\rm live}$ is a factor that penalizes short bursts due to their large trial factors (see e.g., \cite{Braun:2009wp}). In our analysis, we restrict the parameters to $n_s\geq 0, \gamma\in [1,4]\;$ (for power-law fits) $, \sigma_T\in [0,\sigma_{T,{\rm max}}]$, where 
$\sigma_{T,{\rm max}}$ is half the livetime of the dataset. For the combined dataset, we define $\sigma_{T,{\rm max}}$ as half the longest livetime between the datasets.

With the TS defined, we create the background and signal TS distributions by running trials. The background events within a trial are obtained from scrambled data, whereas the signal events are obtained from simulations. We define the 90\% CL sensitivity as the signal flux required for the signal TS to be larger than the background median 90\% of the time. Additionally, we define the $3\sigma$ evidence potential as the injected flux required for the probability of the signal TS to have a $p-$value $p<2.7\times 10^{-3}$ (the $3\sigma$ threshold) is equal to 50\%. The evidence potential at other significances is defined in a similar fashion, by modifying the $p-$value threshold to the corresponding number of $\sigma$.

\begin{figure}
    \centering
    \includegraphics[width=0.8\linewidth]{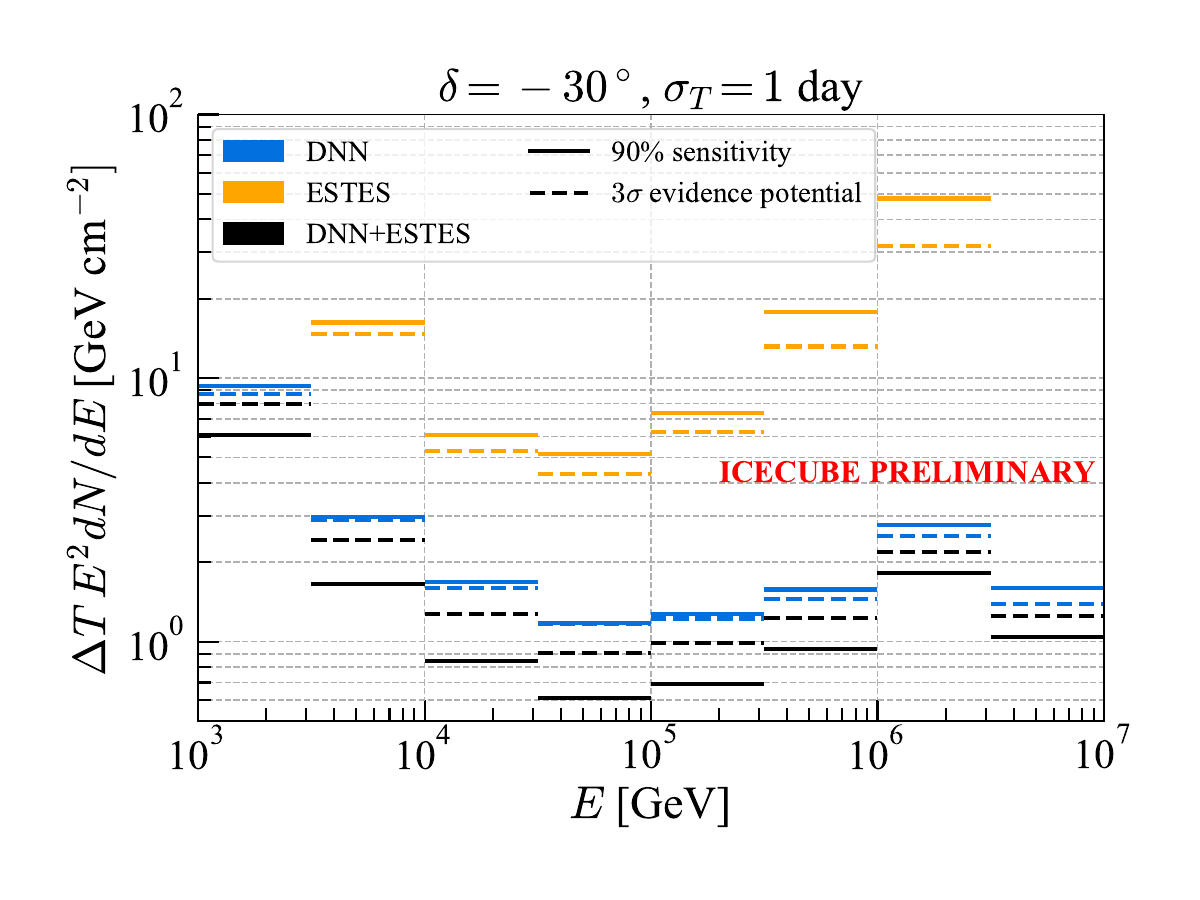}
    \caption{\textcolor{black}{Per-flavor} 90\% CL sensitivity (solid lines) and $3\sigma$ evidence potential (dashed lines) \textcolor{black}{fluences} to a neutrino flare of half-width $\sigma_T$. The blue, orange and black lines correspond to DNNCascade, ESTES and combined datasets, respectively.}
    \label{Figure2}
\end{figure}

\section{Analysis performance}

\subsection{Power-law fits}

In this section, our signal hypothesis to calculate TS will be a point source with an $E^{-\gamma}$ power-law spectrum, \textcolor{black}{where} $\gamma\in[1,4]$, where the fit parameters are $n_s,\gamma,T_0$ and $\sigma_T$.

We show in Figure \ref{Figure1} the $E^{-2}$ sensitivity and evidence potential for different flare half-widths $\sigma_T$, for a source at the horizon. Here we assume $E^{-\gamma}$ signal fluxes, as is typically used in point source searches. We are reporting fluence (time-integrated fluxes) as \textcolor{black}{$\Delta T dN/dE$}, evaluated at 1 TeV.  The left panels show the \textcolor{black}{sensitivity} of the DNNCascade, NT and ESTES samples. Here, the NT data is assuming Gaussian spatial pdfs, not KDEs. The most notable feature is that \textcolor{black}{the} sensitivity worsens for longer-duration flares. For flares shorter than $\sim 0.1$ days, the sensitivity flattens, because we are approaching the background-free regime. Flares longer than 10 days have more background contamination, degrading \textcolor{black}{the} sensitivity. We see that the NT sample provides the best sensitivity at the horizon among the three datasets, as it has the largest statistics and its good angular resolution improves the spatial pdfs for signal events. By combining the datasets, we see a $10\%-15\%$ improvement in sensitivity and evidence potential, but the improvement is less significant for flares longer than $\sim 1$ day. In the right panels of Figure \ref{Figure1}, the NT sample uses KDEs in its signal pdf. This change gives a minor improvement in the sensitivity for \textcolor{black}{flares} shorter than a day. We see that the $3\sigma$ evidence potential is at a lower flux than the sensitivity for short flares, and the two curves cross between $0.1$ and 1 day, depending on the sample. Changing the TS threshold criteria will move the location of the crossing point and is thus dependent on the definition of sensitivity and evidence.

We calculate the differential sensitivity and $3\sigma$ evidence potential for a flare with $\sigma_T=1$ day at a declination $\delta=-30^\circ$. We did this by dividing the energy range into two bins per energy decade, then injected neutrinos at each bin assuming an $E^{-2}$ spectrum. The result is shown in Figure \ref{Figure2}. We see that both the differential sensitivity and $3\sigma$ evidence is best at $E \sim 100$ TeV. Within this energy range, the background event rate is low. As we go to higher energies, we see that the sensitivity worsens. The improvement on the last energy bin is due to the Glashow resonance for cascade events, where the \textcolor{black}{$\bar\nu_e$} charged-current interaction cross section is enhanced. The combined analysis also provides a good improvement when compared to the NT samples alone.

\subsection{Two-sided cutoff fits}
In this section, we introduce an additional energy pdf for the signal hypothesis: an $E^{-2}$ flux with two energy cutoffs of the form

\begin{equation}
    e^{-E_L/E}E^{-2}e^{-E/E_H},
    \label{TwoSidedCutoff}
\end{equation}
where $E_L$ and $E_H$ are the low and high-energy cutoffs, respectively. We will also refer to this flux as a two-sided cutoff. The inclusion of $E_H$ is motivated by the presence of a maximum cosmic ray energy, while $E_L$ accounts for the minimum proton energy required to initiate $p\gamma$ or $pp$ interactions in the source and lead to neutrino production. The use of a more representative signal hypothesis allows for a more accurate sensitivity. Based on the differential sensitivities, having a two-sided cutoff flux that peaks around 100 TeV could potentially be missed in a power-law fit. 

With this new flux hypothesis, the power-law index $\gamma=2$ is fixed when calculating TS. We restrict $E_L\in [100\;{\rm GeV},10\;{\rm TeV}]$ and $E_H\in [10\;{\rm TeV},100\;{\rm PeV}]$. When fitting for two-sided cutoffs, the fit parameters are $n_s,E_L,E_H,T_0$ and $\sigma_T$.

We proceed to compare the performance between flux hypotheses. We assume that the true signal flux is given by Eq. \eqref{TwoSidedCutoff}, with $E_L=1\;{\rm TeV}$ , $E_H=100\;{\rm TeV}$. We then compare the $3\sigma$ evidence potential for this injected flux under a two-sided cutoff and a power-law hypothesis separately. First, we identify the $3\sigma$ evidence potential for the two-sided cutoff hypothesis. We then inject this $3\sigma$ two-sided cutoff flux, as the true signal flux, and perform a power-law fit instead. Note that by swapping to a different signal hypothesis, the signal pdfs going into equation \eqref{TSDefinition} have changed, and thus the TS adopts a different value even if the events in a given trial are the same. Hence, we have to create new signal and background TS distributions. In a power-law fit, the trials for the injected $3\sigma$ two-sided cutoff flux will no longer generate a TS which is 50\% of the time above the $3\sigma$ threshold of the new background TS distribution (i.e. it is not a $3\sigma$ flux under this new hypothesis). Under the power-law hypothesis, we can compute the new number of sigma associated with this two-sided cutoff flux, which we call the recovered significance. In this sense, we say that the $3\sigma$ two-sided cutoff flux has been \textit{recovered} with a different significance.

\begin{figure*}
\includegraphics[width=0.5\textwidth]{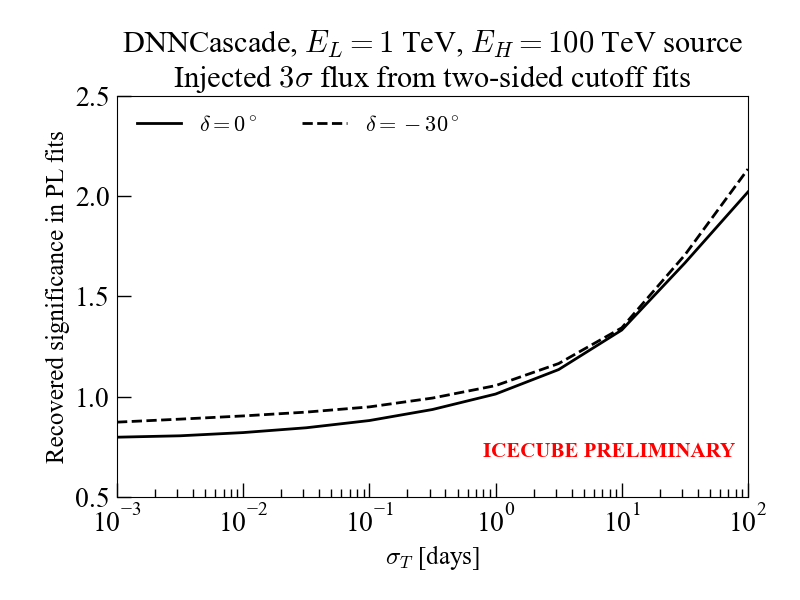}
\includegraphics[width=0.5\textwidth]{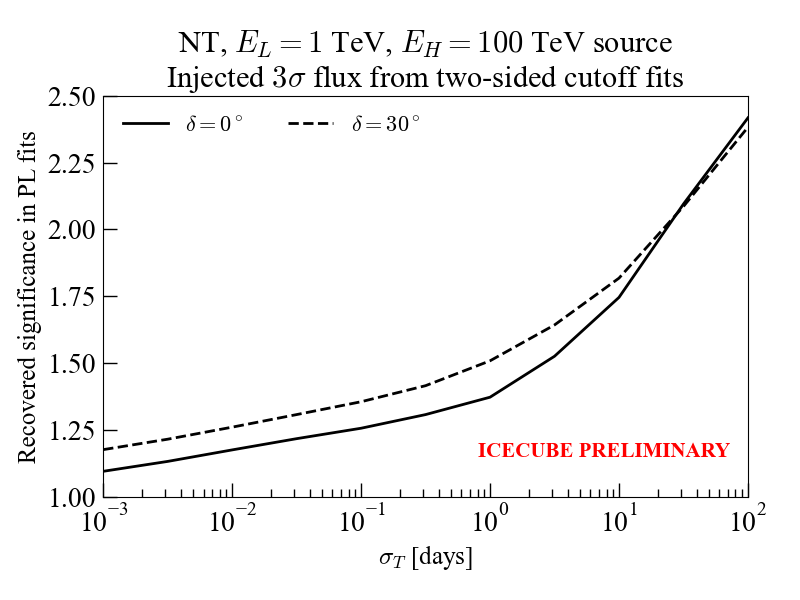}
\caption{Recovered significance in power-law fits after injecting the $3\sigma$ evidence flux for two-sided cutoffs, for different flare half-widths $\sigma_T$. Left (right) panel corresponds to the DNNCascade (NT) sample.}
\label{Figure3}
\end{figure*}

We show the significance recovery in Figure \ref{Figure3}. For this example, the left (right) panel uses the DNNCascade (NT) sample only. The NT sample in the right panel does not use KDEs. We find that a two-sided cutoff flux is not recovered at the $3\sigma$ level in power-law searches. In fact, we only recover it at $1\sigma$ for the shortest time windows. As the flare duration increases, the recovery improves. We found that the TS distribution is not significantly different between signal hypotheses. Hence, the loss in significance is mostly tied to the reconstruction of the signal pdf parameters. The power-law \textcolor{black}{hypothesis} tends to fit for $\gamma>2$, which causes an overfitting of the signal parameter $n_s$. In the case of flares with $\sigma_T\lesssim 1\;{\rm day}$, the injection of a $3\sigma$ two-sided cutoff flux amounts to injecting $\approx 3-4$ signal events. For these values of $\sigma_T$, the injected number of events is too low for the LLH to fully take advantage of the background-free regime and reduce the $n_s$ bias. 

\section{Conclusions}
This is the first multi-flavor time-dependent \textcolor{black}{all-sky} point source search. Here we covered the difference in performance between datasets and its improvement upon combining them. We report a $\approx 15\%$ improvement in sensitivity and evidence potential for $E^{-2}$ sources at the horizon, when using the combined dataset. We also find that a power-law fit in our LLH analysis might miss a two-sided cutoff signal, particularly for flares with $\sigma_T\lesssim 1\;{\rm day}$.

% Bibtex references:
\bibliographystyle{ICRC}
\bibliography{references}

\clearpage

\section*{Full Author List: IceCube Collaboration}

\scriptsize
\noindent
R. Abbasi$^{16}$,
M. Ackermann$^{63}$,
J. Adams$^{17}$,
S. K. Agarwalla$^{39,\: {\rm a}}$,
J. A. Aguilar$^{10}$,
M. Ahlers$^{21}$,
J.M. Alameddine$^{22}$,
S. Ali$^{35}$,
N. M. Amin$^{43}$,
K. Andeen$^{41}$
% ...
\\
\\
% ...
$^{10}$ Universit{\'e} Libre de Bruxelles, Science Faculty CP230, B-1050 Brussels, Belgium \\
% ...
$^{16}$ Department of Physics, Loyola University Chicago, Chicago, IL 60660, USA \\
$^{17}$ Dept. of Physics and Astronomy, University of Canterbury, Private Bag 4800, Christchurch, New Zealand \\
% ...
$^{21}$ Niels Bohr Institute, University of Copenhagen, DK-2100 Copenhagen, Denmark \\
$^{22}$ Dept. of Physics, TU Dortmund University, D-44221 Dortmund, Germany \\
% ...
$^{35}$ Dept. of Physics and Astronomy, University of Kansas, Lawrence, KS 66045, USA \\
% ...
$^{39}$ Dept. of Physics and Wisconsin IceCube Particle Astrophysics Center, University of Wisconsin{\textemdash}Madison, Madison, WI 53706, USA \\
% ...
$^{41}$ Department of Physics, Marquette University, Milwaukee, WI 53201, USA \\
% ...
$^{43}$ Bartol Research Institute and Dept. of Physics and Astronomy, University of Delaware, Newark, DE 19716, USA \\
% ...
$^{63}$ Deutsches Elektronen-Synchrotron DESY, Platanenallee 6, D-15738 Zeuthen, Germany \\
$^{\rm a}$ also at Institute of Physics, Sachivalaya Marg, Sainik School Post, Bhubaneswar 751005, India \\
% ...

\subsection*{Acknowledgments}

\noindent
The authors gratefully acknowledge the support from the following agencies and institutions:
USA {\textendash} U.S. National Science Foundation-Office of Polar Programs,
U.S. National Science Foundation-Physics Division,
U.S. National Science Foundation-EPSCoR,
U.S. National Science Foundation-Office of Advanced Cyberinfrastructure,
Wisconsin Alumni Research Foundation,
Center for High Throughput Computing (CHTC) at the University of Wisconsin{\textendash}Madison,
Open Science Grid (OSG),
Partnership to Advance Throughput Computing (PATh),
Advanced Cyberinfrastructure Coordination Ecosystem: Services {\&} Support (ACCESS),
Frontera and Ranch computing project at the Texas Advanced Computing Center,
U.S. Department of Energy-National Energy Research Scientific Computing Center,
Particle astrophysics research computing center at the University of Maryland,
Institute for Cyber-Enabled Research at Michigan State University,
Astroparticle physics computational facility at Marquette University,
NVIDIA Corporation,
and Google Cloud Platform;
Belgium {\textendash} Funds for Scientific Research (FRS-FNRS and FWO),
FWO Odysseus and Big Science programmes,
and Belgian Federal Science Policy Office (Belspo);
Germany {\textendash} Bundesministerium f{\"u}r Forschung, Technologie und Raumfahrt (BMFTR),
Deutsche Forschungsgemeinschaft (DFG),
Helmholtz Alliance for Astroparticle Physics (HAP),
Initiative and Networking Fund of the Helmholtz Association,
Deutsches Elektronen Synchrotron (DESY),
and High Performance Computing cluster of the RWTH Aachen;
Sweden {\textendash} Swedish Research Council,
Swedish Polar Research Secretariat,
Swedish National Infrastructure for Computing (SNIC),
and Knut and Alice Wallenberg Foundation;
European Union {\textendash} EGI Advanced Computing for research;
Australia {\textendash} Australian Research Council;
Canada {\textendash} Natural Sciences and Engineering Research Council of Canada,
Calcul Qu{\'e}bec, Compute Ontario, Canada Foundation for Innovation, WestGrid, and Digital Research Alliance of Canada;
Denmark {\textendash} Villum Fonden, Carlsberg Foundation, and European Commission;
New Zealand {\textendash} Marsden Fund;
Japan {\textendash} Japan Society for Promotion of Science (JSPS)
and Institute for Global Prominent Research (IGPR) of Chiba University;
Korea {\textendash} National Research Foundation of Korea (NRF);
Switzerland {\textendash} Swiss National Science Foundation (SNSF).

\end{document}